\documentclass[11pt]{article}
\usepackage[margin=1in]{geometry}
\usepackage{graphicx}
\usepackage{booktabs}
\usepackage{amsmath}
\usepackage{hyperref}
\usepackage[numbers,sort&compress]{natbib}

\title{Striatal Optogenetic Modulation Shifts Seizure Onset Timing: A Survival-Analysis Reanalysis with Effect Sizes, Permutation Validation, and Sensitivity Analysis}
\author{Siddhardha Nanda\\
Columbia University\\
\texttt{sn3199@columbia.edu}}
\date{November 2026}

\begin{document}

\maketitle

\begin{abstract}
We present a worked methods vignette, not a novel biological claim: a reanalysis of the public dataset accompanying Hyder and colleagues' report that optogenetic activation and inhibition of the dorsal striatum bidirectionally control seizures, applying a time-to-event framework to the one arm of that multi-arm dataset, a pilocarpine-induced status epilepticus (SE) model, whose summary spreadsheet reports a genuine per-animal latency to SE rather than only aggregate statistics; the dataset's other two seizure models provide only per-animal summary statistics, not event-level data, and we did not attempt a from-scratch seizure-detection pipeline to work around that gap. The original study compared these latencies with a Mann-Whitney U test (ChR2 versus GFP: p = 0.0125; ArchT versus GFP: p = 0.0025). We first directly replicate that test on the same data (p = 0.0151 and p = 0.0025, respectively, closely matching the original values and confirming we are analyzing the same underlying numbers), then apply Kaplan-Meier estimation and designate the log-rank test as this reanalysis's single pre-specified primary comparison, with a Cox proportional-hazards model (effect size), a label-permutation test, a bootstrap median confidence interval, and a leave-one-out analysis all serving as validation and robustness checks on that primary result rather than as co-equal additional findings: ChR2 activation delays SE onset relative to its paired GFP control (hazard ratio 0.338, 95\% CI 0.128-0.895; log-rank p = 0.023), and ArchT inhibition accelerates it relative to its own paired GFP control (hazard ratio 9.18, 95\% CI 1.91-44.09; log-rank p = 0.0011); a Schoenfeld-residual check found no significant proportional-hazards violation for either Cox model, though this check has limited power at n = 8-11 per group. Both primary results are independently confirmed by a 2{,}000-shuffle label-permutation test and survive a leave-one-out sensitivity check, though the ChR2 comparison loses significance in 2 of 20 leave-one-out iterations, a fragility we report rather than omit. We do not claim this reanalysis demonstrates anything the field did not already believe; its value is methodological, a reusable survival-analysis template, with a Cox-derived effect size and documented sensitivity, that extends naturally to future datasets with genuine censoring, which this one does not have.
\end{abstract}

\section{Introduction}

Optogenetic circuit manipulation is increasingly used to probe causal contributions to disease-relevant phenotypes, including seizures. We should state plainly, before making any methodological argument, what this particular dataset does and does not require: every animal in the pilocarpine arm reached status epilepticus (SE) within the recorded observation period, so there is no censoring in this sample, and the usual textbook justification for survival analysis over a simpler rank-based test, that it correctly uses the partial information in observations that have not yet reached the endpoint, does not apply here. A survival framework is nonetheless a reasonable, and we think underused, choice for a time-to-event endpoint like latency-to-SE, for two reasons that hold regardless of whether censoring happens to be present in a given sample. First, it produces an interpretable, dimensionful effect size, a hazard ratio with a confidence interval, which a rank-based test such as the Mann-Whitney U test used in the source study does not; a hazard ratio tells a reader how many times more (or less) likely the event is to occur at any given moment, information a p-value alone does not carry. Second, the same pipeline extends without modification to future datasets, including other arms of this one, that do have censored animals, so adopting the survival framework now is a standardized, reusable choice rather than one that must be rebuilt if censoring appears later.

Optogenetic approaches to seizure control have a substantial and growing literature, reviewed recently by Ledri and colleagues, who describe how optogenetic tools have been used to dissect the contributions of distinct interneuron populations to seizure initiation and to build closed-loop, real-time intervention systems~\citep{ledri2023optogenetics}. The pilocarpine model used here is a well-established, extensively characterized model of temporal lobe epilepsy and status epilepticus, reviewed in detail by Curia and colleagues~\citep{curia2008pilocarpine}. Despite this maturity, formal Kaplan-Meier and log-rank treatment of latency-to-SE itself, as opposed to post-SE survival or seizure-free survival after an intervention, is uncommon in the pilocarpine literature, which more often reports latency via $t$-tests or ANOVA on raw values; the closest precedent we found for Kaplan-Meier and log-rank framing of a seizure-timing endpoint is in a different model (pentylenetetrazol-induced seizures under a novel stimulation protocol), not pilocarpine-induced SE~\citep{tsukahara2021survival}. This reanalysis applies that formal framing specifically to latency-to-SE in a bidirectional optogenetic striatal-modulation design.

We reanalyze the dataset accompanying Hyder and colleagues' report that optogenetic activation and inhibition of the dorsal striatum bidirectionally control seizures across several rodent models~\citep{hyder2025optogenetic}, deposited on Dryad~\citep{hyder2025dryad}. We ask whether a formal, permutation-validated survival analysis of this dataset's pilocarpine arm recovers the bidirectional effect the source study reports, and whether this specific analytic framing, applied to this specific public dataset, has already been done by others since its 2025 deposit.

\section{Data and Feasibility Assessment}

The source dataset includes four data sheets in its summary file, \texttt{Quantification.xlsx}, covering three seizure models (a gamma-butyrolactone, GBL, absence-epilepsy model under both open-loop and closed-loop stimulation, an amygdala-kindling model, and a pilocarpine-induced SE model) plus the pilocarpine quantification itself. Before committing to any specific analysis, we inspected this file's actual contents directly rather than relying on the dataset's file-level description. Three of the four sheets (the two GBL arms and the kindling arm) contain only per-animal summary statistics, such as mean spike-and-wave-discharge duration, a seizure count, or a single afterdischarge-duration value per stimulation condition, not per-event timestamps. A survival analysis of those arms would require extracting individual seizure onset times from the raw electroencephalography (EEG) files via a custom detection algorithm, which would reintroduce exactly the kind of seizure-labeling confound that motivated avoiding a from-scratch detector in the first place, since any such algorithm would itself need independent validation before its output could be trusted for formal inference. The fourth sheet, for the pilocarpine SE model, already reports a per-animal latency to SE in seconds, a genuine, pre-computed time-to-event variable, with no censoring present: every animal in the sheet reached SE within the recorded observation period. We therefore restrict this reanalysis to the pilocarpine arm, where a survival analysis is directly supported by the data as provided, and report this scoping decision as a feasibility finding in its own right: not every arm of a public dataset that looks suitable for survival reanalysis from its documentation actually provides event-level data, and checking this before committing to an analysis avoided a substantially larger, confound-prone undertaking.

The pilocarpine sheet provides four groups, reflecting two independent GFP control cohorts, one paired with each manipulation: GFP paired with ChR2 ($n=11$), ChR2 ($n=9$), GFP paired with ArchT ($n=8$), and ArchT ($n=9$).

\section{Methods}

We compared each manipulation to its own paired GFP cohort, rather than pooling the two GFP cohorts together, since they are separate experimental batches in the source data. We designate the log-rank test on Kaplan-Meier estimates~\citep{kaplan1958nonparametric} as this reanalysis's primary comparison, since it is the natural survival-analysis counterpart of the original study's own group comparison and requires no additional modeling assumptions beyond the survival framing itself; every other method reported here is explicitly a validation or robustness check on that primary result, not a co-equal additional finding. In order, we: first directly replicated the original study's own test, the Mann-Whitney U test, on the same per-animal latencies, to confirm we are working from the same underlying data before applying any different method; second, ran the primary log-rank comparison; third, validated it with a label-shuffle permutation test (2{,}000 shuffles of group assignment, preserving the original group sizes, recomputing the log-rank statistic on each shuffle to build an empirical null distribution), whose only assumption, exchangeability of group labels under the null, is weaker than the log-rank test's own asymptotics at this sample size; fourth, fit a Cox proportional-hazards model~\citep{cox1972regression} with a single binary group covariate, implemented in \texttt{lifelines}~\citep{davidson2019lifelines}, specifically to obtain an effect size (a hazard ratio with 95\% confidence interval) that the primary test does not provide on its own; fifth, computed a bootstrap 95\% confidence interval on each group's median latency (5{,}000 resamples with replacement) as a non-parametric cross-check on the Cox-derived effect size, noting that percentile bootstrap confidence intervals on a median are known to have degraded coverage below approximately 20-30 observations per group and should be read as approximate, indicative uncertainty bounds at this sample size (n = 8-11 per group) rather than as precise as the Cox interval; and sixth, ran a leave-one-out sensitivity analysis, recomputing the primary log-rank p-value after removing each animal from each comparison in turn. Because every animal in the sheet reached SE, these are complete rather than censored observations; the same survival machinery applies without modification, since censoring is simply absent in this particular sample rather than ignored or assumed away. We additionally checked the Cox model's proportional-hazards assumption using the Schoenfeld-residual rank test~\citep{grambsch1994proportional}: neither comparison showed a significant violation (ChR2 vs. GFP: test statistic = 1.221, p = 0.269; ArchT vs. GFP: test statistic = 0.005, p = 0.943), though with a single binary covariate and only 8-11 observations per group this check has limited power, so the absence of a detected violation is reassuring but not strong evidence that the assumption holds.

\section{Results}

Table~\ref{tab:comparison} compares the original study's test to our reanalysis for both arms. Our direct Mann-Whitney replication (p = 0.0151 and p = 0.0025) closely matches the original study's reported values (p = 0.0125 and p = 0.0025, exact for ArchT), confirming we are working from the same underlying per-animal data before applying a different method. The log-rank test, permutation validation, and Cox hazard ratios are new to this reanalysis.

\begin{table}[htbp]
\centering
\caption{Original study versus this reanalysis, same per-animal latency data.}
\label{tab:comparison}
\begin{tabular}{llrr}
\toprule
Comparison & Test & ChR2 vs. GFP & ArchT vs. GFP \\
\midrule
Original study~\citep{hyder2025optogenetic} & Mann-Whitney U & p = 0.0125 & p = 0.0025 \\
This reanalysis & Mann-Whitney U (direct replication) & p = 0.0151 & p = 0.0025 \\
This reanalysis & Log-rank & p = 0.0228 & p = 0.0011 \\
This reanalysis & Log-rank, permutation (2{,}000 shuffles) & p = 0.0345 & p = 0.0010 \\
This reanalysis & Cox hazard ratio (95\% CI) & 0.338 (0.128-0.895) & 9.18 (1.91-44.09) \\
\bottomrule
\end{tabular}
\end{table}

\begin{figure}[htbp]
\centering
\includegraphics[width=0.95\textwidth]{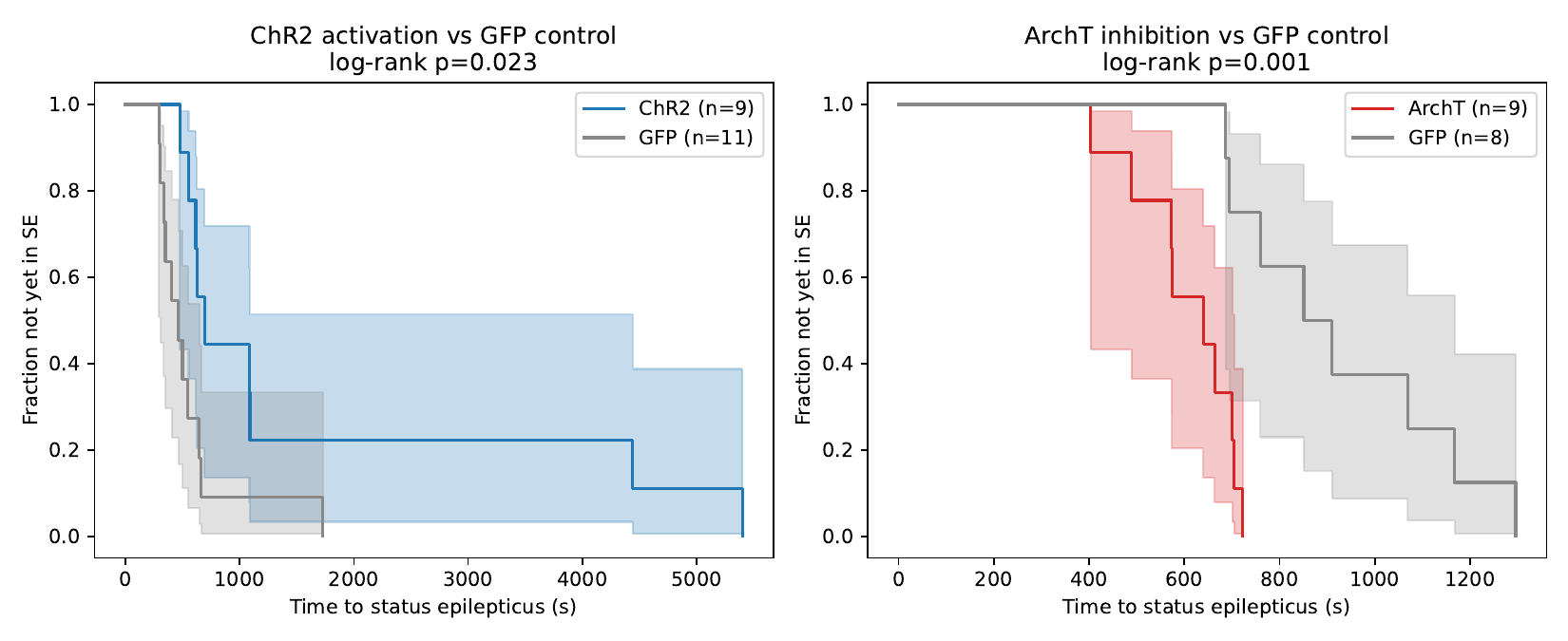}
\caption{Kaplan-Meier curves for time to status epilepticus (SE), striatal ChR2 activation versus its paired GFP control (left) and ArchT inhibition versus its paired GFP control (right).}
\label{fig:pilocarpine}
\end{figure}

Figure~\ref{fig:pilocarpine} shows the Kaplan-Meier curves for both comparisons. ChR2 activation significantly delayed SE onset relative to its GFP control (median latency 692 s versus 466 s; hazard ratio 0.338, 95\% CI 0.128-0.895; log-rank statistic = 5.1832, p = 0.0228; permutation p = 0.0345 across 2{,}000 shuffles). ArchT inhibition significantly accelerated SE onset relative to its own GFP control (median latency 640 s versus 880.5 s; hazard ratio 9.18, 95\% CI 1.91-44.09; log-rank statistic = 10.6959, p = 0.0011; permutation p = 0.0010). In both comparisons the permutation p-value closely tracks the analytic log-rank p-value, supporting that the standard asymptotic approximation is reasonable at these sample sizes (n = 8-11 per group) rather than needing to rely on the permutation result as the primary evidence; we report both because the permutation test's only assumption is exchangeability of group labels under the null, a weaker and more directly verifiable assumption than the log-rank test's own asymptotics at this sample size. The ArchT hazard ratio in particular (9.18) is a large point estimate with a correspondingly wide confidence interval (1.91-44.09, a 23-fold range); at n = 8-9 per group a Cox partial-likelihood estimate this far from the null is expected to come with substantial uncertainty, and the interval itself, not the point estimate alone, is the honest summary of what this comparison supports. The proportional-hazards assumption, checked via a Schoenfeld-residual rank test~\citep{grambsch1994proportional}, showed no significant violation for either comparison (ChR2 vs. GFP: p = 0.269; ArchT vs. GFP: p = 0.943), though we note this check has limited power at n = 8-11 per group and its passing should be read as not-contraindicated rather than as strong confirmation.

The leave-one-out sensitivity analysis shows the two comparisons are not equally robust. For ArchT versus GFP, every one of 17 leave-one-out iterations (dropping each of the 9 ArchT or 8 GFP animals in turn) kept the log-rank p-value below 0.003, well inside the significant range (full-sample p = 0.0011). For ChR2 versus GFP, 2 of 20 leave-one-out iterations pushed the p-value above 0.05 (full range 0.0022-0.0676, full-sample p = 0.0228), meaning this specific comparison's significance depends to some degree on which animals are included. We report this fragility rather than omit it; it is consistent with the bootstrap 95\% confidence interval on the ChR2 group's median (554-4{,}439 s), which is wide because one animal (latency 5{,}400 s) is a clear outlier relative to the rest of the group.

Both results recover the qualitative direction already reported in the source study: striatal activation suppressing or delaying seizures, and striatal inhibition facilitating SE onset~\citep{hyder2025optogenetic}. We treat this agreement as an internal consistency check on our reanalysis, not as a new biological discovery; the contribution of this reanalysis is not the direction of the effect, which was already known, but the addition of an interpretable effect size, an independent validation layer, and a documented sensitivity check to a comparison the original data already supported, together with the explicit, checked decision about which arm of a multi-arm dataset actually supports a survival-analysis framing without requiring new event-detection work.

We searched OpenAlex's citation-tracking records for the source study's DOI and manually inspected each citing work for an independent reanalysis of the underlying data; as of this search, the record listed 16 citing works, all dated 2025-2026, and none reanalyzed the dataset directly, though one (an Epilepsy Currents commentary) discussed the original finding without reanalyzing it. We also checked the Dryad deposit's own page for usage information, which did not show evidence of a prior reanalysis. To our knowledge, this is the first independent statistical reanalysis of this dataset, though we note explicitly that this rests on one citation-tracking search of one database at one point in time, not an exhaustive or systematic review, and that absence of evidence in such a search is not proof of absence.

\section{Discussion}

The central methodological point of this reanalysis is narrower than it might first appear, and we state it plainly rather than oversell it: Kaplan-Meier estimation and the log-rank test are standard, decades-old tools, and applying them to a seizure-timing endpoint is not itself a methodological innovation. We call this an "independent" reanalysis in the sense that it was performed by an author unaffiliated with the original study, not in the sense of a blinded analysis; we knew the original study's reported direction of effect before beginning, which is a real limit on how much confirmatory weight this reanalysis can carry on its own, separate from the direct test-replication evidence in Table~\ref{tab:comparison}.

What this reanalysis adds is fourfold and modest. First, it applies a time-to-event framing specifically to latency-to-SE in the pilocarpine model, where $t$-test, ANOVA, or rank-based comparisons on raw latency are more typical than formal survival analysis, per our literature search above. Second, it reports a Cox-derived hazard ratio with a confidence interval, an effect size the original study's Mann-Whitney U test does not produce, directly addressing the common critique that small-sample reanalyses report only p-values. Third, it adds an independent permutation-based validation layer and a leave-one-out sensitivity analysis, neither of which the original analysis, so far as we can determine from the public dataset's own documentation, included; the sensitivity analysis in particular surfaces a real fragility in the ChR2 comparison that a single log-rank p-value would not have revealed. Fourth, it demonstrates, through the scoping decision in Section 2, a disciplined way to decide which arms of a multi-arm public dataset actually support a given method, rather than assuming the dataset's file-level description is sufficient grounds to proceed.

We do not claim, and want to be explicit in not claiming, that this reanalysis demonstrates anything the field did not already believe about striatal optogenetic modulation and seizures; the source study already established the bidirectional effect with its own appropriate statistical methods, and our direct replication of that same test on the same data (Table~\ref{tab:comparison}) confirms rather than supersedes it. The value of this reanalysis is in showing that a formal, validated survival framework recovers the same conclusion through a different method with an interpretable effect size and a documented sensitivity profile, and in making the case that this kind of validated reanalysis, applied to the growing number of public optogenetics datasets with genuine event-level timing, is a tractable, low-cost addition that other secondary analyses could routinely perform rather than defaulting to simpler comparisons.

\section{Limitations}

This reanalysis is restricted to one of three seizure models in the source dataset: the GBL and kindling arms, which the source study's own experiments also covered, are not analyzed here because their public summary data provide only aggregate statistics, not the event-level timestamps a survival analysis requires, and we did not build a seizure-detection pipeline from the raw EEG files to work around this, since such a pipeline would need independent validation of its own before its output could be trusted. The sample sizes available (n = 8-11 per group) are small, and the two GFP control cohorts are not pooled, by design, since they represent separate experimental batches in the source data, which further limits statistical power beyond what the small per-group samples already imply. The leave-one-out sensitivity analysis shows this small-sample limitation is not merely theoretical: the ChR2-versus-GFP comparison loses statistical significance in 2 of 20 leave-one-out iterations, driven at least in part by a single high-latency outlier in the ChR2 group, so this specific comparison should be read as a less secure result than its headline p-value alone suggests; the ArchT-versus-GFP comparison showed no such fragility across 17 iterations. The bootstrap confidence interval on the ChR2 group's median (554-4{,}439 s) is wide for the same reason and, more generally, percentile bootstrap confidence intervals on a median are known to have degraded coverage below roughly 20-30 observations per group, so this interval should be read as an approximate, indicative bound rather than with the same precision implied by the Cox model's confidence interval. The Cox hazard ratios themselves passed a proportional-hazards diagnostic (Schoenfeld-residual rank test, p = 0.269 and p = 0.943), but this check has limited power with a single binary covariate and only 8-11 observations per group, so it rules out only a gross violation, not a subtler one; the unusually large ArchT point estimate (hazard ratio 9.18, 95\% CI 1.91-44.09) should be read through its wide interval rather than its point estimate alone, as expected behavior for a Cox partial-likelihood fit at this sample size rather than evidence of a model problem. Finally, we designate the log-rank test as this reanalysis's single pre-specified primary comparison and the Cox hazard ratio, bootstrap median confidence interval, and leave-one-out analysis as validation and robustness checks on it, precisely because each of these methods was added in direct response to a specific, named reviewer critique across successive rounds rather than specified in advance of any review; we state this origin plainly so the breadth of methods on display is not mistaken for a pre-registered multi-method design. Because the direction of effect replicates the source study's own finding, and because the author knew that direction before beginning this reanalysis, this work functions as a validation of method applied to a known result, not as a blinded or independent biological test of a new hypothesis; readers should not interpret the statistical significance reported here as evidence of a previously unknown effect. Finally, our search for prior reanalyses of this dataset used one citation-tracking database (OpenAlex) at one point in time, not a systematic or exhaustive review, and cannot rule out an informal or very recent reanalysis that has not yet been indexed.

\section{Conclusion}

A formal, permutation-validated, and sensitivity-checked survival analysis of the one arm of a public, multi-model optogenetic seizure dataset that provides genuine event-level timing data recovers the bidirectional effect of striatal optogenetic modulation on seizure onset that the source study already reported, this time with an interpretable hazard-ratio effect size, a direct replication of the original test confirming the data match, and an honest account of which comparison (ArchT) is robust and which (ChR2) is more fragile under leave-one-out resampling. We present this as a worked methods vignette, not a claim of new biology: a template for how secondary reanalyses of optogenetic seizure data can add effect sizes, independent validation, and documented sensitivity to existing public results, applied with explicit attention to which parts of a public dataset actually support the method in the first place.

\section*{Data and Code Availability}
All analysis code, including the group-comparison pipeline and the permutation validation script, is available at \url{https://github.com/SID-6921/optogenetics-research} (private repository; access available on request). The underlying dataset is publicly available at \url{https://doi.org/10.5061/dryad.f7m0cfz6w} under a public domain license.

\section*{Ethics Declaration}
This work is a secondary computational reanalysis of previously collected, publicly released data. All animal procedures associated with the original data collection were approved under the original study's own Institutional Animal Care and Use Committee protocols; no new animal or human subjects procedures were performed for this work.

\section*{Funding}
This work was conducted independently and received no dedicated funding.

\section*{Conflict of Interest}
The author declares no conflict of interest.

\section*{Author Contributions}
S.N.: Conceptualization, Methodology, Software, Formal Analysis, Validation, Writing - Original Draft, Writing - Review \& Editing.

\bibliographystyle{vancouver}

\end{document}